\documentclass[sigconf]{acmart}

\AtBeginDocument{%
  }

\setcopyright{acmlicensed}
\copyrightyear{2018}
\acmYear{2018}
\acmDOI{XXXXXXX.XXXXXXX}
\acmConference[WWW ’25 Companion]{}{28 April--2 May, 2025}{Sydney, Australia}
\acmISBN{978-1-4503-XXXX-X/18/06}

\usepackage{amsmath,amsfonts}
\usepackage{graphicx}
\usepackage{textcomp}
\usepackage{xcolor}
\usepackage{listings}
\usepackage{enumitem}
\usepackage{microtype}
\usepackage{caption}
\usepackage{subcaption}
\usepackage{booktabs}
\usepackage[ruled,vlined,linesnumbered]{algorithm2e}
\usepackage{setspace}
\usepackage[most]{tcolorbox}
\usepackage{bbding}
\usepackage{multirow}
\usepackage{multicol}
\usepackage{colortbl}

\usepackage{xspace}

\begin{document}

\title{\dataset: A Large-scale Benchmark for Software Engineering Question Answering}
\author{Ruida Hu}
\email{200111107@stu.hit.edu.cn}
\affiliation{
    \institution{Haribin Institute of Technology, Shenzhen}
    \city{Shenzhen}
    \country{China}
}

\author{Chao Peng}
\email{pengchao.x@bytedance.com}
\affiliation{
    \institution{ByteDance}
    \city{Beijing}
    \country{China}
}

\author{Jingyi Ren}
\email{jingyi.422@bytedance.com}
\affiliation{
    \institution{ByteDance}
    \city{Shenzhen}
    \country{China}
}

\author{Bo Jiang}
\email{jiangbo.jacob@bytedance.com}
\affiliation{
    \institution{ByteDance}
    \city{Shenzhen}
    \country{China}
}

\author{Xiangxin Meng}
\email{mengxiangxin.1219@bytedance.com}
\affiliation{
    \institution{ByteDance}
    \city{Beijing}
    \country{China}
}

\author{Qinyun Wu}
\email{wuqinyun@bytedance.com}
\affiliation{
    \institution{ByteDance}
    \city{Beijing}
    \country{China}
}

\author{Pengfei Gao}
\email{gaopengfei.se@bytedance.com}
\affiliation{
    \institution{ByteDance}
    \city{Beijing}
    \country{China}
}

\author{Xinchen Wang}
\email{200111115@stu.hit.edu.cn}
\affiliation{
    \institution{Haribin Institute of Technology, Shenzhen}
    \city{Shenzhen}
    \country{China}
}

\author{Cuiyun Gao}
\email{gaocuiyun@hit.edu.cn}
\affiliation{
    \institution{Haribin Institute of Technology, Shenzhen}
    \city{Shenzhen}
    \country{China}
}

\renewcommand{\shortauthors}{Hu et al.}
\newcommand{\dataset}{CodeRepoQA\xspace}
\newcommand{\wxc}[1]{\textcolor{purple}{{#1}}}
\begin{abstract}

In this work, we introduce \textbf{\dataset}, a large-scale benchmark specifically designed for evaluating repository-level question-answering capabilities in the field of software engineering.
\dataset encompasses five programming languages and covers a wide range of scenarios, enabling comprehensive evaluation of language models.
To construct this dataset, we crawl data from 30 well-known repositories in GitHub, the largest platform for hosting and collaborating on code, and carefully filter raw data.
In total, \dataset is a multi-turn question-answering benchmark with 585,687 entries, covering a diverse array of software engineering scenarios, with an average of 6.62 dialogue turns per entry.

We evaluate ten popular large language models on our dataset and provide in-depth analysis. We find that LLMs still have limitations in question-answering capabilities in the field of software engineering, and medium-length contexts are more conducive to LLMs' performance. The entire benchmark is publicly available at \url{https://github.com/kinesiatricssxilm14/CodeRepoQA}.

\end{abstract}




\maketitle

\section{Introduction}
Large language models (LLMs) are increasingly being integrated into tools such as chatbots and coding assistants to assist developers, showcasing their potential to solving various software engineering tasks~\cite{hou2023large, liu2024large}. As a result, the research community has begun exploring how LLMs can be further leveraged to assist with more complex repository-level tasks encountered in software development~\cite{fan2023automated, macneil2023experiences,ross2023programmer, yang2024swe, liu2024marscode}. GitHub issues provide rich, real-world data that includes bug reports, feature requests, and usage questions, making them an ideal source for constructing diverse and realistic question-answering tasks for evaluating LLMs. Their collaborative nature and technical content allow models to be assessed on practical software development challenges.

Existing question-answering (QA) benchmarks, such as MMLU~\cite{hendryckstest2021}, evaluate models in zero-shot and few-shot settings and assess them in answering question across 57 subjects spanning science, technology and mathematics. CodeQA~\cite{liu2021codeqa} contains a benchmark of 119,778 Java and 70,085 Python QA pairs. CS1QA~\cite{lee2022cs1qa} contains 9,237 question-answer pairs extracted from chat logs of introductory Python classes. CodeApex~\cite{fu2023codeapex} evaluates LLMs on C++ code generation and correction.

Additionally, these benchmarks do not focus on repository-level question-answering and cannot reflect the complexity of real-world scenarios.
Specifically, developers often engage in multi-turn dialogues to resolve issues. Multi-turn dialogues better reflect real-world conversations, as solving software engineering issues often requires multiple interactions~\cite{mtbench2024}.
These benchmarks are limited to single-turn dialogues, which fail to capture the multi-turn interactions needed to resolve software engineering issues. The comparison between the \dataset and other benchmarks can be found in Table \ref{tab:data_compare}.


\definecolor{darkgreen}{rgb}{0,0.5,0}
\newcommand{\cross}{\textcolor{red}{\textbf{\XSolidBrush}}}
\newcommand{\tick}{\textcolor{darkgreen}{\Checkmark}}

\begin{table}[!htbp]
    \caption{The comparison between existing benchmarks and \dataset.}
    \fontsize{7.2}{11}\selectfont
    \centering
    \setlength{\tabcolsep}{0.7mm}
    \begin{tabular}{c|c|c|c|c}
    \toprule
    \textbf{Benchmark} & \textbf{\#Langs} & \textbf{\#Samples} & \textbf{Multi-turn} & \textbf{Repo-level}
\\
\midrule
MMLU~\cite{hendryckstest2021}
& -
& 15,908
& \cross
& \cross
\\

CodeQA~\cite{liu2021codeqa}
& 2
& 190,000
& \cross
& \cross
\\

CS1QA~\cite{lee2022cs1qa}
& 1
& 9,237
& \cross
& \cross
\\

CodeApex~\cite{fu2023codeapex}
& 1
& 250
& \cross
& \cross
\\


\cellcolor{blue!10}\textbf{\dataset}
& \cellcolor{blue!10}{\textbf{5}}
& \cellcolor{blue!10}{\textbf{585,687}}
& \cellcolor{blue!10}{\tick}
& \cellcolor{blue!10}{\tick}
\\ 
    \bottomrule
    \end{tabular}
    \label{tab:data_compare}
\end{table}


To address the above challenges in existing benchmarks, we present \dataset, a novel large-scale benchmark derived from conversations in GitHub repository issues. The benchmark contains 585,687 entries, with an average of 6.62 dialogue turns per conversation. \dataset encompasses five commonly used programming languages, with data details provided in Table \ref{tab:statistics}.

We also conduct extensive experiments with state-of-the-art models in the \dataset, including GPT, DeepSeek-Coder, and the Gemini series, 
and assess the accuracy of the model predictions against ground truth, using metrics such as BLEU, ROUGE-L, ROUGE-1, and Edit Similarity. We find that LLMs still demonstrate limitations in real-world QA scenarios, and medium-length contexts are more conducive to LLMs’ performance.

In summary, the paper makes the following contributions:
\begin{enumerate}
    \item \textbf{Large-scale repository-level QA benchmark from GitHub.} We build a large-scale QA benchmark derived from real-world GitHub repositories. This benchmark includes 585,687 QA entries in five programming languages and encompasses a wide variety of tasks that reflect the complexities of actual software development and maintenance, offering a more realistic assessment of LLM capabilities.
    \item \textbf{Multi-turn software engineering dialogue.} Our benchmark includes multi-turn dialogues, derived from real developer feedback during the development process, with an average of 6.62 dialogue turns per entry.

    \item \textbf{Comprehensive experimental results.} We evaluate state-of-the-art LLMs on the QA scenario. Our experiments reveal key insights into the effectiveness of these models, highlighting their limitations in the QA capabilities within the field of software engineering and the fact that medium-length contexts are more conducive to LLMs' performance.
    \definecolor{darkgreen}{rgb}{0,0.5,0}

\begin{table}[t]
    \caption{Statistics of each programming language in \dataset, where \#average-turn indicates the average turns of the dialogues.}
    \footnotesize
    \renewcommand{\arraystretch}{0.9}
    \centering
    \setlength{\tabcolsep}{1.3mm}
    \begin{tabular}{c|ccccc|c}
    \toprule
    \textbf{Languages} & \textbf{Python} & \textbf{Java} & \textbf{Typescript} & \textbf{Javascript} & \textbf{Go} & \textbf{All}
\\
\midrule
\#entries
& 204,501
& 34,792
& 220,884
& 39,246
& 86,264
& 585,687
\\
\#ave-turn
& 6.61
& 4.31
& 5.89
& 6.98
& 9.32
& 6.62
\\
    \bottomrule
    \end{tabular}
    \label{tab:statistics}
\end{table}
\end{enumerate}

\section{Data Construction}
\begin{figure}[t]
	\centering
	\includegraphics[width=0.48\textwidth]{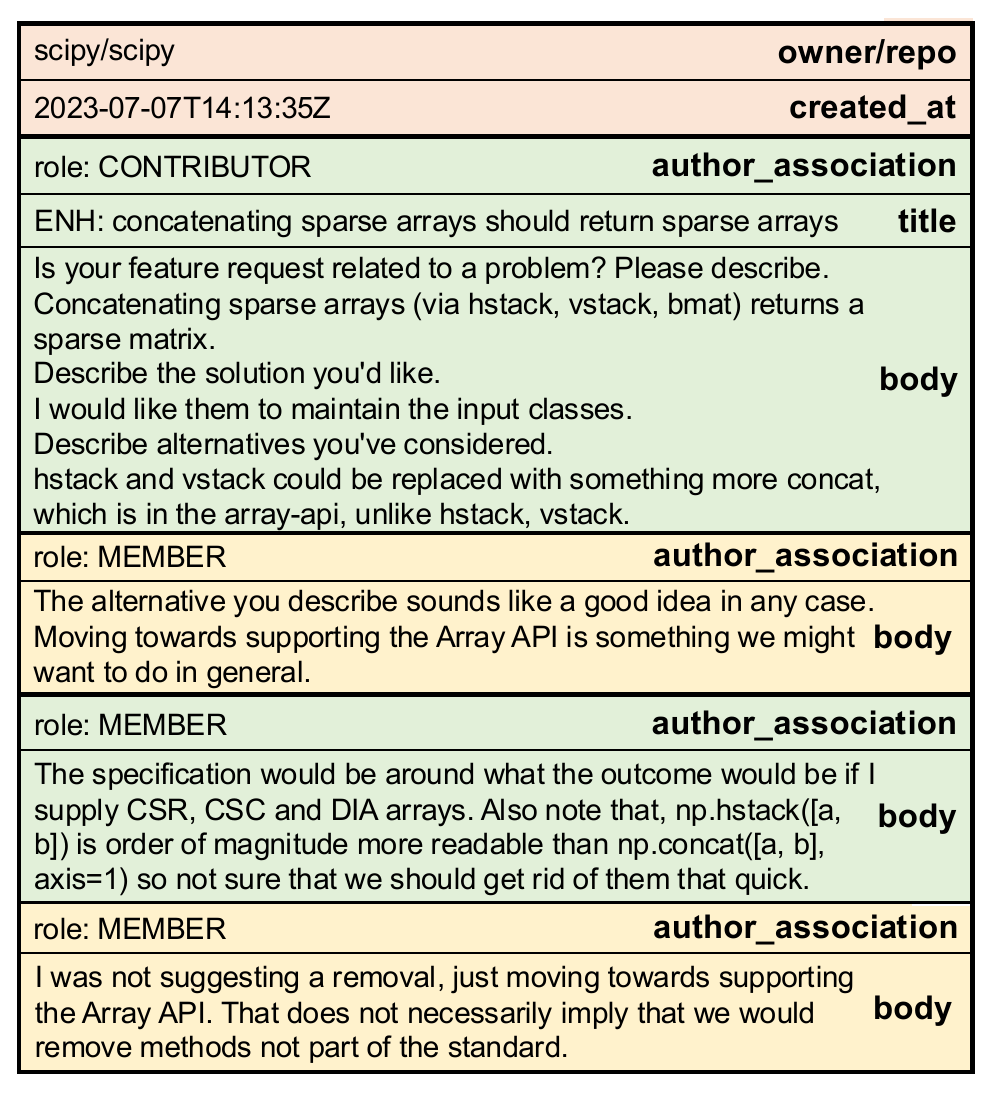}
    \caption{A multi-turn QA entry illustrating the main components of the entry in \dataset.}
\label{fig:structure}
\end{figure}

We employ two steps to produce the final benchmark\footnote{The crawling was performed in August 2024, and the complete structure is displayed in our GitHub repository.}: 
\textbf{(1) Raw Data Crawling}: we crawl and extract metadata related
to issues from GitHub;
\textbf{(2) Data Filtering}: we select and clean the gathered data to ensure its relevance and quality.
After completing the above two steps, we obtain the final benchmark with diverse issue attributes.
As shown in Figure \ref{fig:structure}, the data includes key attributes such as repository information (owner, repo,  time) and main QA details (issue title, issue body, and responses). Additionally, the role indicates the user's relationship to the repository, such as contributor or member.

\subsection{Raw Data Crawling}
To ensure sufficient diversity within \dataset, we select 30 popular repositories from GitHub, encompassing five widely-used programming languages: Python, Java, JavaScript, TypeScript, and Go. To assure the quality of the benchmark, we select repositories with over 5,000 stars that are widely utilized within their respective fields. GitHub, as the largest platform for hosting and managing software projects, contains a wealth of metadata. The GitHub REST API~\cite{restapi} allows developers to interact with GitHub services through HTTP requests. We use the GitHub REST API to crawl all issues from 30 repositories on GitHub, totalling over 636,000 entries. To ensure the quality of \dataset, we conducted the following filtering.

\subsection{Data Filtering}
After completing the Raw Data Crawling process, we obtain approximately 636,000 issues. These issues are intended for constructing the QA benchmark. We apply a filtering process to ensure high-quality data automatically. We apply the following criteria to filter the issue entries:

\textbf{Comment scale}: Prioritize character count; remove entries under 200 characters or those exceeding 10MB to balance sufficient information and model input limits.

\textbf{Redundancy}: According to the GitHub documentation~\cite{duplicatedocument}, we exclude duplicate issues by detecting markers such as "\texttt{Duplicate of \#}" to ensure unique and valuable entries.

\textbf{External check}: Filter out conversations that contain external links, and only retain issues containing internal GitHub links to avoid complications from external content.

\textbf{Event count}: Influenced by methodologies from research like StarCoder\cite{starcoder}, we exclude issues with more than ten events to avoid auto-generated text and robot-produced records, maintaining high-quality data.

\textbf{Robot detection}: Identify and remove issues and responses created by robots through inspection of the author information and some common robot response patterns.

\textbf{Participant count}: Exclude issues with only one participant, after removing robot users, to ensure the benchmark contains genuine and effective interactions.

After completing the data filtering process, we obtained 585,687 issue entries containing numerous real-world code-related QA pairs. Our benchmark includes five programming languages; the number of entries and dialogue turns for each language are detailed in Table \ref{tab:statistics}.

\section{Experiment}
The following experiments use filtered QA pairs as the benchmark.
For our constructed benchmark, \dataset, we propose the following two Research Questions (RQs) based on the specified aspects:
\begin{enumerate}[label=\bfseries RQ\arabic*:,leftmargin=.5in]
    \item How do models perform in answering questions?
    \item How does the length of question affect the models' answering performance?
\end{enumerate}

\subsection{Model Selection}
\label{3_1}
We selected ten different large language models (LLMs) to evaluate their performance on coding tasks. These models are broadly used and represent some of the most advanced technologies available, including both commercial and open-source options.
The commercial models include the GPT and Gemini (GM) series, such as GPT-4o, GPT-4, Gemini-1.5-Flash, and Gemini-1.5-Pro. The open-source models include Mistral, CodeQwen (CQ), and the DeepSeek Coder (DSC) family, encompassing a range of model sizes. These include Mistral-large-2 (123B), CodeQwen-1.5-Chat (7B), DeepSeek-Coder-V2 (236B), DeepSeek-Coder-V2-Lite (16B), DeepSeek-Coder (33B), and DeepSeek-Coder (6.7B).
To ensure the validity of all experiments, we strictly controlled the length of the input data to ensure it did not exceed the maximum input tokens allowed by the selected models.

\subsection{Evaluation Design}
\label{sec:taskdesign}
To comprehensively assess the capability of LLMs in question-answering in software engineering scenarios, we designed a specific QA task. For this evaluation, we use the historical dialogue turns as input, ensuring that the model understands the context of the conversation. The last response from a repository maintainer (such as MEMBER, AUTHOR, or CONTRIBUTOR) is used as the ground truth. This approach allows us to accurately answer the LLMs' ability to answer the questions in software engineering scenarios.




\subsection{Evaluation Metrics}
We employ the following metrics to evaluate LLMs, including BLEU, ROUGE-L, ROUGE-1, and Edit Similarity (ES), all of which range from 0 to 1.

\textbf{BLEU}: Measures the n-gram precision between the generated text and reference texts, incorporating a brevity penalty to discourage overly concise outputs.

\textbf{ROUGE-L and ROUGE-1}: These metrics assess the quality of text generated by models compared to reference texts. ROUGE-L evaluates sequence-level similarity using the Longest Common Subsequence (LCS) method, while ROUGE-1 measures the overlap of single words (1-grams). The F1 score is the primary metric for a balanced evaluation of content accuracy and completeness.

\textbf{Edit Similarity}: Evaluates the similarity between the generated text and the reference text by comparing the number of edits required to transform one text into the other.



\subsection{Performance of LLMs}
\begin{table*}[!htbp]
    \caption{Performance of various LLMs on \dataset. In the table, color shades of each block denote performance rankings: darkest for highest, medium for second highest, and lightest for third highest scores.}
    \fontsize{7.5}{11}\selectfont
    \renewcommand{\arraystretch}{0.9}
    \centering
    \setlength{\tabcolsep}{1.0mm}
    \setlength{\extrarowheight}{1.1mm}
    \begin{tabular}{c|cccc|cccccc|c}
    \toprule
    \textbf{Metric} & \textbf{GPT-4o} &\textbf{GPT-4} & \textbf{GM-Flash} & \textbf{GM-Pro} & \textbf{Mistral-123B} & \textbf{CQ-7B} & \textbf{DSC-236B} & \textbf{DSC-16B} & \textbf{DSC-33B} & \textbf{DSC-6.7B} & \textbf{Average}
\\
\midrule
BLEU &
0.0943 &
0.1179 &
\cellcolor{black!40}{0.1227} &
\cellcolor{black!20}{0.1208} &
0.0948 &
\cellcolor{black!10}{0.1188} &
0.1102 &
0.0942 &
0.1184 &
0.1182 &
0.1110
\\

ROUGE-L &
0.1189 &
0.1330 &
\cellcolor{black!20}{0.1499} &
\cellcolor{black!40}{0.1551} &
0.1228 &
0.1392 &
0.1340 &
0.1201 &
\cellcolor{black!10}{0.1401} &
0.1366 &
0.1350
\\

ROUGE-1 &
0.1984 &
\cellcolor{black!10}{0.2315} &
\cellcolor{black!20}{0.2410} &
\cellcolor{black!40}{0.2470} &
0.1988 &
0.2264 &
0.2203 &
0.1961 &
0.2272 &
0.2279 &
0.2215
\\

Edit Similarity &
0.1388 &
0.1715 &
\cellcolor{black!20}{0.1914} &
\cellcolor{black!40}{0.2074} &
0.1411 &
\cellcolor{black!10}{0.1803} &
0.1615 &
0.1406 &
0.1785 &
0.1784 &
0.1689
\\

\hline
Average &
0.1376 &
0.1635 &
\cellcolor{black!20}{0.1762} &
\cellcolor{black!40}{0.1826} &
0.1393 &
\cellcolor{black!10}{0.1662} &
0.1565 &
0.1377 &
0.1660 &
0.1653 &
0.1591
\\





    \bottomrule
    \end{tabular}
    \label{tab:rq1}
\end{table*}
We evaluate the performance of ten models listed in Section \ref{3_1}. The results for the evaluation are presented in the respective sections of Table \ref{tab:rq1}. We can derive the following observations:


\textbf{Commercial models do not always outperform open-source ones in software engineering, nor do larger models consistently show superior performance.} Although the two models from the Gemini series achieved the best overall performance, GPT-4, a famous commercial model, underperforms compared to open-source models like CodeQwen (7B). Additionally, the largest and newest model in the DeepSeek series, DeepSeek-Coder-V2, is outperformed by its smaller counterparts, 33B and 6.7B. This suggests that having more parameters does not necessarily lead to better performance, highlighting that larger models are not always superior across different tasks.


\textbf{LLMs still demonstrate limitations in software engineering QA scenarios.} Even with Gemini-1.5-Pro, which has the best overall performance, there is a significant gap between its output and the actual feedback. The combined score across four metrics is only 0.1826, with a BLEU score of 0.1208. Specifically, Gemini-1.5-Pro achieved the highest scores in ROUGE-L (0.1551) and ROUGE-1 (0.2470). This indicates that the vocabulary overlap between the generated text and real-world responses is relatively low, suggesting that the generated content may not effectively capture key information from the actual responses.


\begin{table}[!htbp]
    \caption{The impact of different context lengths on the performance of model responses. The dark blocks indicate the group with the best performance for the metric. The $\sim$ symbol followed by a percentage indicates the proportion of each group's length.}
    \fontsize{7.5}{11}\selectfont
    \renewcommand{\arraystretch}{0.9}
    \centering
    \setlength{\tabcolsep}{1.2mm}
    \begin{tabular}{c|ccccc|c}
    \toprule
    \textbf{Metric} & \textbf{$\sim$20\%} &\textbf{$\sim$40\%} & \textbf{$\sim$60\%}  & \textbf{$\sim$80\%} & \textbf{$\sim$100\%} & \textbf{Average}
\\
\midrule
BLEU &
0.1077 &
\cellcolor{black!40}{0.1196} &
0.1107 &
0.1092 &
0.1080 &
0.1110
\\

ROUGE-L &
0.1393 &
\cellcolor{black!40}{0.1440} &
0.1373 &
0.1312 &
0.1230 &
0.1350
\\

ROUGE-1 &
0.2240 &
\cellcolor{black!40}{0.2348} &
0.2212 &
0.2132 &
0.2140 &
0.2215
\\

Edit Similarity &
0.1708 &
\cellcolor{black!40}{0.1806} &
0.1664 &
0.1648 &
0.1621 &
0.1689
\\

\midrule
Average &
0.1605 &
\cellcolor{black!40}{0.1698} &
0.1589 &
0.1546 &
0.1518 &
0.1591
\\





    \bottomrule
    \end{tabular}
    \label{tab:rq2}
\end{table}
\subsection{Impact of Question Length on LLMs' Performance}
To investigate the impact of question length on performance of LLMs, we equally divide the benchmark into five groups based on the question length. We then calculate the scores for each group across four metrics. The detailed results are shown in Table \ref{tab:rq2}.

\textbf{LLMs perform better on questions of medium length in software engineering QA scenarios.} Based on the data in Table \ref{tab:rq2}, it is evident that LLMs perform better on questions of medium length. The 40\% context length group achieves the highest scores in BLEU (0.1196), ROUGE-1 (0.2348), and Edit Similarity (0.1806), and maintains the highest average score (0.1698) among all groups. Similarly, the 60\% context length group also performs well, particularly in BLEU, ROUGE-L, and ROUGE-1 metrics, with an average score of 0.1589. In contrast, the shortest (20\%) and longest (100\%) context length groups show relatively lower performance across most metrics, indicating that extremely short and long contexts are less effective. These findings suggest that medium-length contexts are more conducive to generating high-quality answers from LLMs in software engineering QA scenarios.

\section{Conclusion}
In this paper, we present \dataset, a large-scale benchmark for assessing the question-answering capabilities of LLMs in the field of software engineering. \dataset is a multi-turn question-answering benchmark, containing 585,687 entries from 30 well-known GitHub repositories and covers five distinct programming languages.
\dataset provides a comprehensive evaluation of LLMs’ abilities in answering questions of repository-level compared to previous benchmarks.
Our experiments show that LLMs still demonstrate limitations in software engineering QA scenarios, and medium-length questions are found to yield better performance in generating high-quality answers.
Our benchmark and detailed description are available at: \url{https://github.com/kinesiatricssxilm14/CodeRepoQA}.

\bibliographystyle{ACM-Reference-Format}
\bibliography{references}

\end{document}